\documentclass[reprint,superscriptaddress,showpacs,amsmath,amssymb,aps,pra]{revtex4-2}
\usepackage{pifont}
\usepackage{times}
\usepackage{amssymb}
\usepackage{graphicx}
\usepackage{dcolumn}
\usepackage{dsfont}
\usepackage{bm}
\usepackage{subfigure}
\usepackage[ruled]{algorithm2e}
\usepackage{hyperref}
\usepackage{appendix}
\usepackage{lipsum}

\hypersetup{colorlinks=true, citecolor=blue, urlcolor=blue, linkcolor=blue}

\newtheorem{theorem}{Theorem}

\newtheorem{example}{Example}
\newtheorem{lemma}{Lemma}

\input amssym.def

\begin{document}
\title{Parameterized coherence measure}
\author{Meng-Li Guo}
\email{guoml@cnu.edu.cn}
\affiliation{School of Mathematical Sciences, Capital Normal University, Beijing 100048, China}
\author{Zhi-Xiang Jin}
\email{jzxjinzhixiang@126.com}
\affiliation{School of Computer Science and Technology, Dongguan University of Technology, Dongguan, 523808, China}
\author{Jin-Min Liang}
\affiliation{School of Mathematical Sciences, Capital Normal University, Beijing 100048, China}
\author{Bo Li}
\email{libobeijing2008@163.com}
\affiliation{School of Computer and Computing Science, Hangzhou City University, Hangzhou 310015, China}
\author{Shao-Ming Fei}
\email{feishm@cnu.edu.cn}
\affiliation{School of Mathematical Sciences, Capital Normal University, Beijing 100048, China}
	
\begin{abstract}
Quantifying coherence is an essential endeavor for both quantum mechanical foundations and quantum technologies. We present a bona fide measure of quantum coherence by utilizing the Tsallis relative operator $(\alpha, \beta)$-entropy. We first prove that the proposed coherence measure fulfills all the criteria of a well defined coherence measure, including the strong monotonicity in the resource theories of quantum coherence. We then study the ordering of the Tsallis relative operator $(\alpha, \beta)$-entropy of coherence, Tsallis relative $\alpha$-entropies of coherence, R\'{e}nyi $\alpha$-entropy of coherence and $l_{1}$ norm of coherence for both pure and mixed qubit states.
This provides a new method for defining new coherence measure and entanglement measure, and also provides a new idea for further study of quantum coherence.
\end{abstract}
	
\maketitle
	
\section{Introduction}
Coherence is a fundamental aspect of quantum physics. It is an important resource in quantum information processing \cite{Nielsen} and plays significant roles in the emerging fields such as quantum metrology \cite{Lloyd,Dobrzanski2014}, nanoscale thermodynamic \cite{Aberg,Lostaglio} and quantum biology \cite{Sarovar,Lloyd1,Huelga,Lambert}. The quantification of coherence has attracted much attention recently \cite{Gour,Marvian,Levi,Plenio,Aberg1,Yu,Guo2020}. To quantify the coherence Baumgratz, Cramer and Plenio established a consistent framework in terms of quantum resource theory \cite{Plenio}. Fruitful results have been obtained in characterizing quantum coherence both theoretically and experimentally \cite{Plenio,Fan,c10,cq1,c11,X2,JX,c12}.
In particular, the distillable coherence \cite{c12}, the coherence of formation \cite{yzcm}, the robustness of coherence \cite{nbcp}, the coherence measures based on entanglement \cite{auhm}, the max-relative entropy of coherence \cite{bukf}, the R\'{e}nyi entropy of coherence\cite{Renyi1,Renyi2}, the Tsallis relative entropies of coherence\cite{Tallis} and the coherence concurrence \cite{dbq} have been proposed and investigated.
With the development of quantum coherence theory, the basic properties of different coherence measures have been well studied, and important work showing that quantum coherence has important physical significance has also been obtained one after another\cite{QJ1,QJ2,QJ3}. So it is of great significance to find more coherence measures.

Moreover, many important properties of quantum coherence measures have been also explored. Among them is the quantum state ordering of quantum coherence measures. A given physical resource may have different quantitative measures. It is of importance whether two different quantum states have the same ordering under different measures. Taking the entanglement resource theory as an example, the ordering problem of the concurrence and the negativity has been discussed in \cite{Ord}. It is found that these entanglement measures may give rise to different orderings. More discussions about the entanglement ordering are given in \cite{Ei99,Vi00,Zy02,We03,Zi06,Se03}.
Concerning quantum coherence, if two different coherence measures $C_1$ and $C_2$ satisfy that $C_1 (\rho_1)\leq C_1 (\rho_2) \Leftrightarrow C_2(\rho_1)\leq C_2(\rho_2)$ for any quantum states $\rho_1$ and $\rho_2$, it is said that $C_1$ and $C_2$ have the same quantum state ordering. Liu \cite{Liu16} proves that the relative entropy of coherence and $l_1$-norm coherence have the same quantum state ordering for any single qubit pure states, but not for general high-dimensional quantum states or single-qubit mixed states.
More works has been done toward the quantum state ordering related to coherence measures \cite{MCD,Zhang}.

Recently, Nakamura and Umegaki extended the notion of the von Neumann entropy and presented the operator entropy $-A\log A$ for any positive operator $A$ on a Hilbert space $\mathcal{H}$ \cite{Nak92}. The relative entropy $A\log A-A\log B$ for positive operators $A$ and $B$ is also introduced associated with the semifinite von Neumann algebra. Moreover, Fujii and Kamei \cite{SGH33} introduced the relative operator entropy for two reversible positive operators $A$ and $B$. This concept is a generalization of operator entropy and relative entropy. In addition, Furuta \cite{SGH39} also got the parametric generalization of Shannon inequality. For two-person reversible positive operator $A$ and $B$ and arbitrary real number $\alpha \in (0,1]$ in Hilbert space, K. Yanagi et al. introduced the concept of Tsallis relative operator $(\alpha, \beta)$-entropy $T_{\alpha, \beta}(A|B)$ \cite{SGH113,Glasgow2019}. The authors in Ref \cite{SGH36,SGH38} studied the properties of the Tsallis relative operator entropy and obtained the generalization of the Shannon's inequality that plays an important role in classical information theory, e.g., its operator form can be applied to quantum thermo dynamics and quantum information theory.

In this work, we propose a well-defined coherence measures via the Tsallis relative operator $(\alpha, \beta)$-entropy. Perspective mapping has been widely used in the research of quantum information theory and quantum statistical mechanics, and the Tsallis relative operator $(\alpha, \beta)$-entropy is the perspective mapping of the function $\ln_{\alpha} x:=\frac{x^{\alpha}-1}{\alpha}$\cite{SGH28,SGH57,SGH77,SGH78}. Therefore, our results can provide a new method for defining new entanglement measure and coherence measure. That is, through perspective mapping of existing measure functions, entanglement or coherence measure with similar or better properties can be obtained. Finally, we also study the quantum state ordering of coherence measure, and compare the new parameterized coherence measures with the $l_1$-norm coherence, Tsallis relative $\alpha$-entropies of coherence and R\'{e}nyi $\alpha$-entropy of coherence. It is proved that they have the same quantum state ordering under certain parameters.

\section{Coherence via Tsallis relative operator $(\alpha, \beta)$-entropy}
Let $\mathcal{H}$ be a $d$-dimensional Hilbert space with orthogonal basis $\{|i\rangle\}^d_{i=1}$. With respect to this basis, the set of the incoherent states $\mathcal{I}$ has the form $\delta=\sum_{i=1}^d\delta_i|i\rangle\langle i|$,
where $\delta_i\in[0,1]$ and $\sum_i\delta_i=1$.
Any proper measure of coherence $C$ should satisfy the following axioms \cite{Plenio}:
(C1) $C(\rho)\geq0$ for all quantum states $\rho$, and $C(\rho)=0$ if and only if $\rho\in \mathcal{I}$;
(C2) Monotonicity under incoherent completely positive and trace preserving maps (ICPTP) $\Psi$, $C(\rho) \geq C(\Psi(\rho))$;
(C3) Monotonicity for average coherence under subselection based on measurements outcomes: $C(\rho)\geq\sum_i p_iC(\rho_i )$, where $\rho_i= K_i\rho K_i^\dagger/p_i$, $p_i=\mathrm{Tr}( K_i\rho K_i^\dagger)$ for all ${K_i}$ satisfying $\sum_iK_i^\dagger K_i=I$ ($I$ denotes the identity operator) and $K_i\mathcal{I}K_i^\dagger\subseteq \mathcal{I}$;
(C4) Non-increasing under mixing of quantum states (convexity), i.e., $\sum_ip_iC(\rho_i)\geq C(\sum_ip_i\rho_i)$ for any ensemble $\{p_i, \rho_i\}$.
Note that conditions (C3) and (C4) automatically imply condition (C2). The condition (C3) allows for the sub-selection of measurement outcomes in well controlled experiments.

The $l_{1}$ norm of coherence~\cite{Plenio} is defined by
\begin{equation}\label{Cl1}
C_{l_{1}}(\rho)=\displaystyle\sum_{i\neq j}\mid \rho_{ij}\mid,
\end{equation}
where $\rho_{ij}$ are entries of $\rho$. The
Tsallis relative $\alpha$-entropies of coherence $C^{T}_{\alpha_1}(\rho)$ is defined by \cite{Tallis}
\begin{equation*}
C^{T}_{\alpha_1}(\rho)=\displaystyle\min_{\delta\in I}\frac{1}{\alpha_1-1}\Big [\mathrm{tr}(\rho^{\alpha_1}\delta^{1-\alpha_1})-1 \Big]
\end{equation*}
for $\alpha_1 \in(0,1)\sqcup (1,\infty)$.
The R\'{e}nyi $\alpha$-entropies of coherence is defined by
\begin{equation*}
C^{R}_{\alpha_1}(\rho)=\displaystyle\min_{\delta\in I}\frac{1}{\alpha_1-1}\Big[\log \mathrm{tr}(\rho^{\alpha_1}\delta^{1-\alpha_1}) \Big]
\end{equation*}
for $\alpha_1 \in [0,\infty)$.
$C^{T}_{\alpha_1}(\rho)$ and $C^{R}_{\alpha_1}(\rho)$ have analytic expressions \cite{Tallis},
\begin{eqnarray}
C^{T}_{\alpha_1}(\rho)=\frac{1}{\alpha_1-1} \Big[ \Big( \sum_{i} \langle i| \rho^{\alpha_1} |i \rangle ^{\frac{1}{\alpha_1}} \Big)^{\alpha_1}-1\Big],\label{CT1}
\end{eqnarray}
\begin{eqnarray}
C^{R}_{\alpha_1}(\rho)&=&\frac{\alpha_1}{\alpha_1-1} \log \sum_{i} \langle i| \rho^{\alpha_1} |i \rangle ^{\frac{1}{\alpha_1}}.\label{CR1}
\end{eqnarray}

The Tsallis relative operator $(\alpha, \beta)$-entropy is defined by \cite{Glasgow2019},
\begin{eqnarray}\label{n2}
T_{\alpha, \beta}(\rho||\sigma)=\frac{\rho^{\frac{ \beta}{2}}\left(\rho^{-\frac{ \beta}{2}}\sigma \rho^{-\frac{ \beta}{2}}\right)^{\alpha}\rho^{\frac{ \beta}{2}}-\rho^{ \beta}}{\alpha},
\end{eqnarray}
for arbitrary two invertible positive operators $\rho$ and $\sigma$ on Hilbert space, and any real number $\alpha,\beta \in (0,1]$.
For convenience, writes $T_{\alpha, \beta}(\rho||\sigma)$ as
\begin{eqnarray*}
T_{\alpha, \beta}(\rho||\sigma)=\rho^{\frac{\beta}{2}}\ln_{\alpha}(\rho^{-\frac{\beta}{2}}\sigma \rho^{-\frac{\beta}{2}}) \rho^{\frac{\beta}{2}},
\end{eqnarray*}
where $\ln_{\alpha}X\equiv\frac{X^{\alpha}-1}{\alpha}$ for the positive operator $X=\rho^{-\frac{\beta}{2}}\sigma \rho^{-\frac{\beta}{2}}$.
The perspective mapping $P_f$ of the function $f$ is defined as\cite{SGH57}: $P_f(A,B):=B^{\frac{1}{2}}f(B^{-\frac{1}{2}}AB^{-\frac{1}{2}})B^{\frac{1}{2}}$, where $A$ is a self-adjoint operator on Hilbert space $H$, $B$ is a strictly positive operator and its spectral set falls in a closed interval containing 0. When $A$ and $B$ are noncommutative, Ebadian, et.al, defines an noncommutative generalized perspective map by choosing an appropriate order\cite{SGH77,SGH78}: $P_{f\Delta h}(A,B):=h(B)^{\frac{1}{2}}f(h(B)^{-\frac{1}{2}}Ah(B)^{-\frac{1}{2}})h(B)^{\frac{1}{2}}$.
Obviously, the Tsallis relative operator $(\alpha, \beta)$-entropy is the perspective mapping of the function $\ln_{\alpha} x:=\frac{x^{\alpha}-1}{\alpha}$.
Then we can get that Tsallis relative operator $(\alpha, \beta)$-entropy $T_{\alpha, \beta}(\rho||\sigma)$ has the following properties, see proof in Appendix A.

(T1) (monotonicity) If $\sigma \leq \tau$, then $T_{\alpha, \beta}(\rho||\sigma)\leq T_{\alpha, \beta}(\rho||\tau)$.

(T2) (superadditivity) If $\rho_1 \leq \rho_2$, $\sigma_1 \leq \sigma_2$, then $T_{\alpha, \beta}(\rho_1+\rho_2||\sigma_1+\sigma_2)\geq T_{\alpha, \beta}(\rho_1||\sigma_1)+T_q(\rho_2||\sigma_2)$.

(T3) \cite{Glasgow2019} (joint concavity) $T_{\alpha, \beta}(n \rho_1+(1-n) \rho_2||n\sigma_1+(1-n)\sigma_2)\geq n T_{\alpha, \beta}(\rho_1||\sigma_1)+(1-n) T_{\alpha, \beta}(\rho_2||\sigma_2)$.

(T4) $T_{\alpha, \beta}(U \rho U^{\dagger }||U \sigma U^{\dagger }) =T_{\alpha, \beta}(\rho||\sigma)$ for any unitary operator $U$.

(T5) For a unital positive linear map $\Phi$ from the set of the bounded linear operators on Hilbert space to itself, $\Phi(T_{\alpha, \beta}(\rho||\sigma))\leq T_{\alpha, \beta}(\Phi(\rho)||\Phi(\sigma))$.

Denote
\begin{eqnarray*}
\rho \sharp_{\alpha, \beta} \sigma \equiv \rho^{\frac{\beta}{2}} (\rho^{-\frac{\beta}{2}} \sigma \rho^{-\frac{\beta}{2}})^{\alpha} \rho^{\frac{\beta}{2}}
\end{eqnarray*}
the operator mean between $\rho$ and $\sigma$, so called the $\alpha$-power mean \cite{SGH38}.
Next we have the following lemma about the $\alpha$-power mean.
\begin{lemma}\label{Lm1}
For any quantum states $\rho$ and $\sigma$, and any positive real number $a>0$, $\alpha,\beta \in (0,1]$, the following inequalities hold,
\begin{eqnarray}
&&T_{\alpha, \beta}(\rho || \sigma)\geq \rho \sharp_{\alpha, \beta}\sigma -\frac{1}{a} \rho \sharp_{\alpha-1, \beta} \sigma + (\ln_{\alpha} \frac{1}{a})\rho^{\beta}, \label{lower1}\\
&&T_{\alpha, \beta}(\rho || \sigma) \leq \frac{1}{a} \sigma -\rho^{\beta} - (\ln_{\alpha}\frac{1}{a}) \rho \sharp_{\alpha, \beta} \sigma, \label{up1}
\end{eqnarray}
where $\ln_{\alpha}\frac{1}{a} \equiv \frac{(\frac{1}{a})^{\alpha}-1}{\alpha}$.
Particularly, when $a=1$, the condition $T_{\alpha, \beta}(\rho || \sigma)=0$ is equivalent to $\rho^{\beta}=\sigma$.
\end{lemma}

{\sf Proof}
According to Lemma 3.5 in \cite{SGH38}, we have
\begin{eqnarray}\nonumber
x^{\alpha}\left(1-\frac{1}{a x}\right) + \ln_{\alpha}\frac{1}{a} \leq \ln_{\alpha} x \leq \frac{x}{a} -1 -x^{\alpha}\ln_{\alpha}\frac{1}{a}.
\end{eqnarray}
Set $x=\rho^{-\frac{ \beta}{2}}\sigma \rho^{-\frac{ \beta}{2}}$ in the above formula. We obtain
$$
\begin{array}{rcl}
\ln_{\alpha}(\rho^{-\frac{ \beta}{2}}\sigma \rho^{-\frac{ \beta}{2}}) &\geq & (\rho^{-\frac{ \beta}{2}}\sigma \rho^{-\frac{ \beta}{2}})^{\alpha}\nonumber\\
&& - \frac{1}{a} (\rho^{-\frac{ \beta}{2}}\sigma \rho^{-\frac{ \beta}{2}}) ^{\alpha-1}+ (\ln_{\alpha}\frac{1}{a}) I, \nonumber\\[2mm]
\ln_{\alpha}(\rho^{-\frac{ \beta}{2}}\sigma \rho^{-\frac{ \beta}{2}}) &\leq & \frac{1}{a} (\rho^{-\frac{ \beta}{2}}\sigma \rho^{-\frac{ \beta}{2}}) -I \nonumber\\
&& - (\ln_{\alpha}\frac{1}{a}) (\rho^{-\frac{ \beta}{2}}\sigma \rho^{-\frac{ \beta}{2}})^{\alpha}.\nonumber
\end{array}
$$
Multiplying $\rho^{\frac{ \beta}{2}}$ on both sides of each term in above inequality, we obtain
\begin{eqnarray}
&&\rho^{\frac{ \beta}{2}} \ln_{\alpha}(\rho^{-\frac{ \beta}{2}}\sigma \rho^{-\frac{ \beta}{2}}) \rho^{\frac{ \beta}{2}} \geq \rho \sharp_{\alpha, \beta}\sigma -\frac{1}{a} \rho \sharp_{\alpha-1, \beta} \sigma + (\ln_{\alpha} \frac{1}{a})\rho^{\beta}, \nonumber\\
&&\rho^{\frac{ \beta}{2}} \ln_{\alpha}(\rho^{-\frac{ \beta}{2}}\sigma \rho^{-\frac{ \beta}{2}}) \rho^{\frac{ \beta}{2}} \leq \frac{1}{a} \sigma -\rho^{\beta} - (\ln_{\alpha}\frac{1}{a}) \rho \sharp_{\alpha, \beta} \sigma.\nonumber
\end{eqnarray}
Noting that $\rho^{\frac{ \beta}{2}} \ln_{\alpha}(\rho^{-\frac{ \beta}{2}}\sigma \rho^{-\frac{ \beta}{2}}) \rho^{\frac{ \beta}{2}}=T_{\alpha, \beta}(\rho || \sigma)$, we complete the proof of the inequalities (\ref{lower1}) and (\ref{up1}).

If $T_{\alpha, \beta}(\rho || \sigma)=0$ and $a=1$, from (\ref{lower1}) and (\ref{up1}) we have $\rho^{\beta}\leq \sigma$ and $\sigma \leq \rho^{\beta}$, namely, $\rho^{\beta}=\sigma$. Conversely, if $\rho^{\beta}=\sigma$, it is easily seen that $T_{\alpha, \beta}(\rho || \sigma)=0$. $\Box$

Next, we set $f(\rho, \sigma)=\mathrm{tr}[\rho^{\frac{\beta}{2}}(\rho^{-\frac{\beta}{2}}\sigma \rho^{-\frac{\beta}{2}})^{\alpha}\rho^{\frac{\beta}{2}}]$, where $\alpha \in(0,1]$. The following two lemmas about the function $f(\rho,\sigma)$ is important in deriving our main results. The proofs of Lemmas 2 and 3 are given in Appendix B and C, respectively.

\begin{lemma}\label{lem2}
For any quantum states $\rho$ and $\sigma$ with $supp\,\rho\subseteq supp\,\sigma$, we have $f(\Phi(\rho),\Phi(\sigma))\geq f(\rho,\sigma)$ for any completely positive and trace preserving (CPTP) map $\Phi$.
\end{lemma}

\begin{lemma}\label{SSS} Let $\Phi :=\left\{ K_{n}:\sum_{n}{K}_{n}^{\dagger }{K}_{n}= \mathcal{I}_{H}\right\}$ be a CPTP map which transforms the states $\rho $ and $\sigma $ into the ensembles $\left\{ p_{n},\rho_{n}\right\} $ and $\left\{q_{n},\sigma_{n}\right\} $,
respectively. We have
\begin{equation*}
f\left( \rho _{H},\delta _{H}\right) \leq \sum_{n} p_{n}^{\gamma}q_{n}^{1-\gamma}f_{q}\left( \rho _{n},\sigma _{n}\right)
\end{equation*}
\end{lemma}
for $\gamma\in(0,1)$.

Below we propose a measure of quantum coherence based on the Tsallis relative operator $(\alpha,\beta)$-entropy.

\begin{theorem}\label{Th1} The following parameterized function $C^{T}_{\alpha, \beta}(\rho)$ of state $\rho$ is a bona fide measure of quantum coherence,
\begin{eqnarray}\label{CTab}
C^{T}_{\alpha, \beta}(\rho)=\min_{\sigma \in \mathcal{I}} \frac{1}{\alpha} \{1-[\mathrm{tr}( \rho^{\frac{ \beta}{2}}(\rho^{-\frac{ \beta}{2}}\sigma \rho^{-\frac{ \beta}{2}})^{\alpha} \rho^{\frac{ \beta}{2}})]^{\frac{1}{(1-\alpha)\beta}} \},\nonumber
\end{eqnarray}
where $\alpha \in (0,1)$ and $\beta \in (0,1]$.
\end{theorem}

{\sf Proof }
From Lemma 1, we have $C^T_{\alpha, \beta}(\rho)\geq0$,
and $C^T_{\alpha, \beta}(\rho)=0$ if and only if $\rho=\sigma.$

For any CPTP map $\phi$, by using the property (T5) we get
\begin{eqnarray}
&&\frac{1}{\alpha}\{\text{Tr}(\rho^{\beta})-\text{Tr}[\rho^{\frac{\beta}{2}}(\rho^{-\frac{\beta}{2}}\sigma \rho^{-\frac{\beta}{2}})^{\alpha }\rho^{\frac{\beta}{2}}] \} \nonumber\\
&&\geq \frac{1}{\alpha}\{\text{Tr}(\rho)^{\beta}-\text{tr}[\phi (\rho)^{\frac{\beta}{2}}(\phi (\rho)^{-\frac{\beta}{2}} \phi (\sigma) \phi (\rho)^{-\frac{\beta}{2}})^{\alpha }\phi (\rho)^{\frac{\beta}{2}}] \}. \nonumber
\end{eqnarray}
According to the Jensen's inequality \cite{OP} one gets $\Phi[\rho^{\beta}] \leq \Phi(\rho)^{\beta}$ and
\begin{eqnarray}
&&\frac{1}{\alpha}\{1-\text{tr}[\rho^{\frac{\beta}{2}}(\rho^{-\frac{\beta}{2}}\sigma \rho^{-\frac{\beta}{2}})^{\alpha }\rho^{\frac{\beta}{2}}] \} \nonumber\\
&& \geq \frac{1}{\alpha}\{1-\text{tr}[\phi (\rho)^{\frac{\beta}{2}}(\phi (\rho)^{-\frac{\beta}{2}} \phi (\sigma) \phi (\rho)^{-\frac{\beta}{2}})^{\alpha }\phi (\rho)^{\frac{\beta}{2}}] \}. \nonumber
\end{eqnarray}
This implies that
\begin{eqnarray}
&&\text{tr}[\rho^{\frac{\beta}{2}}(\rho^{-\frac{\beta}{2}}\sigma \rho^{-\frac{\beta}{2}})^{\alpha }\rho^{\frac{\beta}{2}}]\nonumber\\
&&\leq \text{tr}[\phi (\rho)^{\frac{\beta}{2}}(\phi (\rho)^{-\frac{\beta}{2}} \phi (\sigma)\phi (\rho)^{-\frac{\beta}{2}})^{\alpha }\phi (\rho)^{\frac{\beta}{2}}],\nonumber\\
&&\{\text{tr}[\rho^{\frac{\beta}{2}}(\rho^{-\frac{\beta}{2}}\sigma \rho^{-\frac{\beta}{2}})^{\alpha }\rho^{\frac{\beta}{2}}]\} ^{\frac{1}{(1-\alpha)\beta}} \nonumber\\
&&\leq \{\text{tr}[\phi (\rho)^{\frac{\beta}{2}}(\phi (\rho)^{-\frac{\beta}{2}} \phi (\sigma) \phi (\rho)^{-\frac{\beta}{2}})^{\alpha }\phi (\rho)^{\frac{\beta}{2}}]\} ^{\frac{1}{(1-\alpha)\beta}}.\nonumber
\end{eqnarray}
For any ICPTP map $\phi_{ \mathcal{I}},$ there exists $\sigma ^{\ast }\in
\mathcal{I}$ such that
\begin{eqnarray}
&&\max_{\sigma \in \mathcal{I}}\{\text{tr}[\rho^{\frac{\beta}{2}}(\rho^{-\frac{\beta}{2}}\sigma \rho^{-\frac{\beta}{2}})^{\alpha }\rho^{\frac{\beta}{2}}]\} ^{\frac{1}{(1-\alpha)\beta}}  \nonumber \\
&&=\{\text{tr}[\rho^{\frac{\beta}{2}}(\rho^{-\frac{\beta}{2}}\sigma ^{\ast } \rho^{-\frac{\beta}{2}})^{\alpha }\rho^{\frac{\beta}{2}}]\} ^{\frac{1}{(1-\alpha)\beta}}  \nonumber \\
&&\leq \{\text{tr}[\phi _{ \mathcal{I}}(\rho)^{\frac{\beta}{2}}(\phi _{ \mathcal{I}}(\rho)^{-\frac{\beta}{2}}\phi _{ \mathcal{I}}(\sigma ^{\ast }) \phi _{ \mathcal{I}}(\rho)^{-\frac{\beta}{2}})^{\alpha } \phi _{ \mathcal{I}}(\rho)^{\frac{\beta}{2}}]\} ^{\frac{1}{(1-\alpha)\beta}}  \nonumber \\
&&\leq \max_{\sigma \in \mathcal{I}}\{\text{tr}[\phi _{ \mathcal{I}}(\rho)^{\frac{\beta}{2}}(\phi _{ \mathcal{I}}(\rho)^{-\frac{\beta}{2}} \sigma ^{\ast } \phi _{ \mathcal{I}}(\rho)^{-\frac{\beta}{2}})^{\alpha } \phi _{ \mathcal{I}}(\rho)^{\frac{\beta}{2}}]\} ^{\frac{1}{(1-\alpha)\beta}}. \nonumber
\end{eqnarray}
This proves that $C^T_{\alpha, \beta}(\rho)$ satisfies (C2).

That $C^T_{\alpha,\beta}(\rho)$ satisfies (C4) is directly derived from (T3).

To prove that $C^T_{\alpha, \beta}(\rho)$ satisfies (C3),
denote $\delta^{o}$ the optimal incoherent state such that $f(\rho,\delta^{o})=\max_{\delta \in \mathcal{I}} f(\rho,\delta )$.
Let $\Phi= \{K_n\}$ be the incoherent selective quantum operations given by Kraus operators $\{K_n\}$ with $\sum_{n}K_n^{\dagger}K_{n}={I}$.
The operation $\Phi$ on a state $\rho$ gives rise to the post-measurement ensemble $\{p_n, \rho_n\}$ with $p_n = \mathrm{Tr}K_n\rho K_n^{\dagger}$ and $\rho_n=K_n\rho K_n^{\dagger}/p_n$. Hence the averaged coherence is
\begin{equation}\label{pfth11}
\sum_{n}p_n C^T_{\alpha, \beta}(\rho_n)=\min_{\delta _{n}\in \mathcal{I}}\frac{1}{\alpha}\Big[1-\sum_{n}p_{n} f^{\frac{1}{\gamma}} (\rho _{n},\delta _{n})\Big],
\end{equation}
where $\gamma=(1-\alpha)\beta$.
Since the incoherent operation cannot generate coherence from the optimal incoherent state $\delta^{o}$, we have $\delta_n^o=K_n\delta^o K_n^{\dagger}/q_n\in \mathcal{I}$ with $q_n=\mathrm{Tr}K_n\delta^o K_n^{\dagger}$ for any incoherent operator $K_n$.
Since $\alpha \in(0,1)$ and $C^T_{\alpha, \beta}(\rho)\geq 0$, $C^T_{\alpha, \beta}(\rho)$ is the smallest when $f^{\frac{1}{\gamma}}(\rho,\delta)$ is the maximum.
Therefore, one immediately finds that $\mathrm{max}_{\delta \in \mathcal{I}} f^{\frac{1}{\gamma}}(\rho,\delta)\geq f^{\frac{1}{\gamma}}(\rho_n,\delta_n^o)$.
Eq. (\ref{pfth11}) implies then that
\begin{equation}\label{n11}
\sum_{n}p_n C^T_{\alpha, \beta}(\rho_n)\leq\frac{1}{\alpha}\left(1-\sum_{n}p_{n}f^{\frac{1}{\gamma}}\left( \rho _{n},\delta_n^o \right)\right).
\end{equation}

By using the H\"{o}lder inequality
\begin{eqnarray}
\sum_{k=0}^{d}a_k b_k\leq \left(\sum_{k=0}^{d} a^n_k\right)^\frac{1}{n}\left(\sum_{k=0}^{d} b^m_k \right)^\frac{1}{m}\nonumber
\end{eqnarray}
for $ \frac{1}{n}+\frac{1}{m}=1 $ and $ n>1$,
we obtain
\begin{equation}\label{n12}
\left[\sum_{n}q_{n}\right] ^{1-\gamma} \left[\sum_{n}p_{n}f^{\frac{1}{\gamma}}(\rho _{n},\delta _{n}^{o})\right] ^{\gamma}\geq
\sum_{n}p_{n}^{\gamma}q_{n}^{1-\gamma}f (\rho _{n},\delta_{n}^{o}).
\end{equation}
Therefore, (\ref{n11}) becomes
\begin{eqnarray}\label{n13}
\sum_{n}p_n C^T_{\alpha, \beta}(\rho_n) &&\leq \frac{1}{\alpha} \left(1-\sum_{n}p_{n}f^{\frac{1}{\gamma}}( \rho _{n},\delta_n^o)\right)\\\nonumber
&&\leq\frac{1}{\alpha}\left(1-\left[\sum_{n}p_{n}^{\gamma}q_{n}^{1-\gamma}f( \rho_{n},\delta_n^o)\right]^{\frac{1}{\gamma}}\right)\\\nonumber
&&\leq\frac{1}{\alpha}\left(1-f^{\frac{1}{\gamma}}(\rho,\delta^o )\right)\\\nonumber
&&= C^T_{\alpha, \beta}(\rho),\nonumber
\end{eqnarray}
where the first inequality is due to (\ref{n11}), the second inequality is from
(\ref{n12}), the third inequality is due to Lemma \ref{SSS}. (\ref{n13}) shows that $C^T_{\alpha,\beta}(\rho)$ satisfies (C3). $\Box$

Thus, the Tsallis relative operator $(\alpha,\beta)$-entropy of coherence $C^T_{\alpha,\beta}(\rho)$ is a bona fide measure of coherence. Because Tsallis relative operator $(\alpha, \beta)$-entropy is the perspective mapping of the function $\ln_{\alpha} x:=\frac{x^{\alpha}-1}{\alpha}$, and the perspective mapping $P_{f}$ has some good properties, for example, (i) The function $f$ is matrix convex if and only if the perspective function $P_{f}$ is jointly convex, (ii) The function $f$ is matrix concave if and only if the perspective function $P_{f}$ is jointly concave. Therefore, we suspect that we can get a new a new kind of well-defined coherence measure or entanglement measure by perspective mapping it with functions or existing coherence measure. This provides a new method for defining new coherence measure and entanglement measure, and also provides a new idea for us to study quantum coherence.

\section{ordering states with Tsallis relative operator $(\alpha, \beta)$-entropy of coherence}
\subsection{Ordering states with $C^{T}_{\frac{1}{2},1}$, $C^{T}_{\alpha_1}$,  $C^{R}_{\alpha_1}$ and $C_{l_{1}}$ for single-qubit pure states}
In this section, we show that the Tsallis relative operator $(\alpha, \beta)$-entropy, the Tsallis relative $\alpha$-entropies, the R\'{e}nyi $\alpha$-entropy and the $l_{1}$ norm of coherence generate the same ordering for single-qubit pure states.

For simplicity we consider the Tsallis relative operator $(\alpha, \beta)$-entropy of coherence for $\alpha=\frac{1}{2}$ and $\beta=1$. By definition we have
\begin{eqnarray}
C^{T}_{\frac{1}{2},1}(\rho)&=& \min_{\sigma \in \mathcal{I}}2\{1- [ \mathrm{tr}( \rho^{\frac{1}{2}}(\rho^{-\frac{1}{2}}\sigma \rho^{-\frac{ 1}{2}})^{\frac{1}{2}} \rho^{\frac{1}{2}})]^2 \}\nonumber\\
&\geq&\min_{\sigma \in \mathcal{I}}2\Big\{1-\Big[\mathrm{tr}\Big(\rho (\rho^{-\frac{1}{2}}\sigma \rho^{-\frac{1}{2}})\rho\Big)^{\frac{1}{2}}\Big]^2 \Big\}\nonumber\\
&=&\min_{\sigma \in \mathcal{I}}2\Big\{1-[\mathrm{tr}(\rho^{\frac{1}{2}}\sigma \rho^{\frac{1}{2}})^{\frac{1}{2}}]^2 \Big\}, \label{Ctab2}
\end{eqnarray}
where the inequality is due to the Araki-Lieb-Thirring inequality, namely,
for matrices $A, B\geq0$, $s\geq0$ and for $0\leq r\leq1$, it holds that $\mathrm{tr}(A^{r}B^{r}A^{r})^{s}\leq \mathrm{tr}(ABA)^{rs}$ \cite{c32}.

Let $|\psi\rangle=\sqrt{p}|0\rangle + e^{i\varphi}\sqrt{1-p}|1\rangle$ be a general single-qubit pure state, and $\sigma=q|0 \rangle \langle 0|+(1-q)|1 \rangle \langle 1|$ a general incoherent qubit state, $0 \leq p,q \leq 1$. From (\ref{Cl1}), (\ref{CT1}), (\ref{CR1}) and (\ref{Ctab2}), we obtain
\begin{eqnarray}\label{CRT1}
C^{T}_{\frac{1}{2},1}(\rho)&\geq&\min_{q}2[1-(-2 p q+p+q-1)^2],\nonumber\\
C^{T}_{\alpha_1}(\rho)&=&\frac{1}{\alpha_1 -1}\Big[\Big
(p^{\frac{1}{\alpha_1}}+(1-p)^{\frac{1}{\alpha_1}}\Big)^{\alpha_1 }-1\Big],\nonumber\\
C^{R}_{\alpha_1}(\rho)&=&\frac{\alpha_1 }{\alpha_1 -1}\Big[\log \Big(p^{\frac{1}{\alpha_1} }+(1-p)^{\frac{1}{\alpha_1}}\Big)\Big],\nonumber\\
C_{l_1}(\rho)&=&2 \sqrt{p (1-p)}.\nonumber
\end{eqnarray}
Here only $C^{T}_{\frac{1}{2},1}(\rho)$ depends on the incoherent state $\sigma$.
For any given any $p\in [0,1]$, there exists $q^{\ast } \in q$ such that
\begin{eqnarray}
C^{T}_{\frac{1}{2},1}(\rho)
&=&2[1-(-2 p q^{\ast }+p+q^{\ast }-1)^2],
\end{eqnarray}
where $q^{\ast}\in [0,1]$.

\begin{figure}
	\includegraphics[width=7cm]{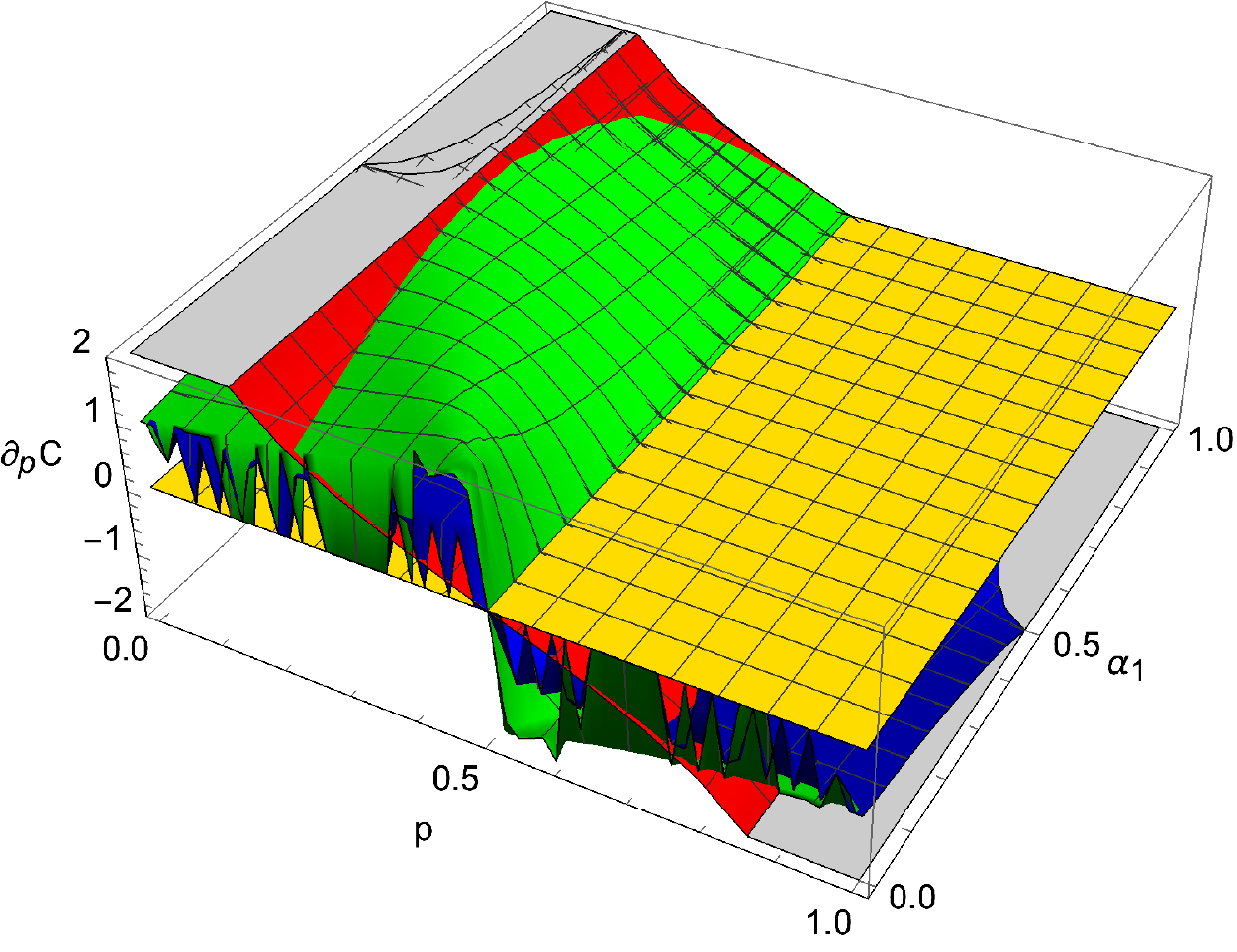}
	\caption{The blue surface is $\frac{\partial C^{T}_{\alpha_1}}{\partial p}$, the green surface is $\frac{\partial C^{R}_{\alpha_1}}{\partial p}$, the red surface is $\frac{\partial C_{l_1}}{\partial p}$, and the yellow surface is the zero plane.}
	\label{Fig1}
\end{figure}

\begin{figure}
	\includegraphics[width=7cm]{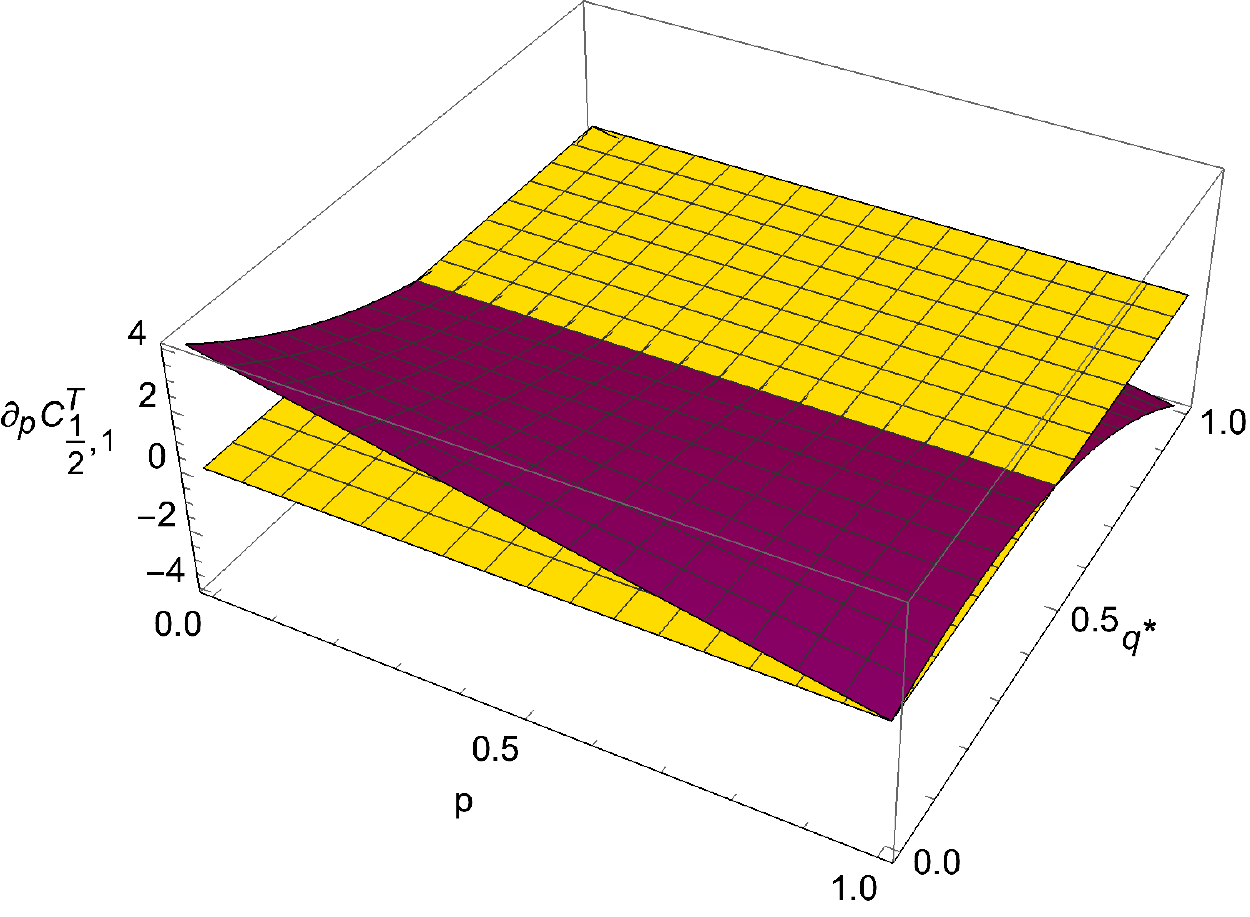}
	\caption{The purple surface is $\frac{\partial C^{T}_{\frac{1}{2},1}}{\partial p}$ and the yellow surface is the zero plane.}
	\label{Figg2}
\end{figure}

From the derivations of the coherence measures with respect to $p$,
\begin{eqnarray}
\frac{\partial C^{T}_{\frac{1}{2},1}}{\partial p}&=&-4 (1-2 q^{\ast }) (-2 p q^{\ast }+p+q^{\ast }-1),\nonumber\\
\frac{\partial C^{T}_{\alpha_1}}{\partial p}&=&\frac{1}{\alpha_1 -1 }\Big[(p^{\frac{1}{\alpha_1 }-1}-(1-p)^{\frac{1}{\alpha_1 }-1}) (p^{\frac{1}{\alpha_1}}\nonumber\\
&&+(1-p)^{\frac{1}{\alpha_1}})^{\alpha_1 -1}\Big],\nonumber\\
\frac{\partial C^{R}_{\alpha_1}}{\partial p}&=&\frac{1}{\alpha_1 -1 }\Big[\frac{p^{\frac{1}{\alpha_1}-1}-(1-p)^{\frac{1}{\alpha_1 }-1}}{p^{\frac{1}{\alpha_1}}+(1-p)^{\frac{1}{\alpha_1}}}\Big],\nonumber\\
\frac{\partial C_{l_1}}{\partial p}&=&\frac{1-2 p}{\sqrt{(1-p) p}},\nonumber
\end{eqnarray}
{\sf Proof}
we have, see Figs. \ref{Fig1} and \ref{Figg2},
\begin{center}
$\frac{\partial C_{l_1}}{\partial p}, \frac{\partial C^{T}_{\alpha_1}}{\partial p},\frac{\partial C^{R}_{\alpha_1}}{\partial p}$
$\left\{
\begin{array}
	{cc}> 0,& \quad  0<p<\frac{1}{2}, 0<\alpha_1<\frac{1}{2},\\
	<0, &\quad  \frac{1}{2}<p<1,  0<\alpha_1<\frac{1}{2}\\
	>0,& \quad  0<p<\frac{1}{2}, \frac{1}{2}<\alpha_1<1,\\
	<0,& \quad  \frac{1}{2}<p<1,  \frac{1}{2}<\alpha_1<1,\\
\end{array}
\right. $
and
$\frac{\partial C^{T}_{\frac{1}{2},1}}{\partial p}$
$\left\{
\begin{array}
	{cc}> 0,& \quad  0<p<\frac{1}{2}, 0<q^{\ast }<\frac{1}{2},\\
	<0, &\quad  0<p<\frac{1}{2}, \frac{1}{2}<q^{\ast }<1,\\
	>0,& \quad  \frac{1}{2}<p<1, 0<q^{\ast }<\frac{1}{2},\\
	<0,& \quad  \frac{1}{2}<p<1, \frac{1}{2}<q^{\ast }<1.\\
\end{array}
\right.$
\end{center}
Therefore, we obtain that $(1)$. $C_{l_{1}}$ is an increasing function for $p\leq \frac{1}{2}$, and it is a decreasing function for $p\geq \frac{1}{2}$;
$(2)$. For any $0<\alpha_1 <1$, $C^{T}_{\alpha_1}$, $C^{R}_{\alpha_1}$ are increasing function when $0\leq p\leq \frac{1}{2}$, they are decreasing function with $\frac{1}{2}\leq p\leq 1$;
$(3)$. For any $0\leq p\leq 1$, $C^{T}_{\frac{1}{2},1}$ is an increasing function when $0\leq q^{\ast } \leq \frac{1}{2}$, and it is a decreasing function for $\frac{1}{2} \leq q^{\ast } \leq 1$.

From the above analysis, we see that $C^{T}_{\frac{1}{2},1}$, $C^{T}_{\alpha_1}$,  $C^{R}_{\alpha_1}$ and $C_{l_{1}}$ have the same monotonicity for single qubit pure states under the specific conditions. Without loss of generality, set $0\leq p_1,p_2 \leq \frac{1}{2}$ and $0 \leq q \leq \frac{1}{2}$. Taking into account the above properties (1)-(3), we have $C^{T}_{\frac{1}{2},1}(| \psi\rangle)\leq C^{T}_{\frac{1}{2},1}( |\varphi\rangle)$ if and only if $p_1\leq p_2$, $C^{T}_{\alpha_1}(| \psi\rangle)\leq C^{T}_{\alpha_1}( |\varphi\rangle)$ if and only if $p_1\leq p_2$, $C^{R}_{\alpha_1}(| \psi\rangle)\leq C^{R}_{\alpha_1}( |\varphi\rangle)$ if and only if $p_1\leq p_2$, and $p_1\leq p_2$ if and only if $C_{l_1}(| \psi\rangle)\leq C_{l_1}( |\varphi\rangle)$. Therefore, we have the following theorem.

\begin{theorem}\label{Th2}
For any two single-qubit pure states $|\psi\rangle=\sqrt{p_1}|0\rangle + \sqrt{1-p_1}|1\rangle$ and $|\varphi\rangle=\sqrt{p_2}|0\rangle + \sqrt{1-p_2}|1\rangle$, and $\sigma=q|0 \rangle \langle 0|+(1-q)|1 \rangle \langle 1|$, $0 \leq p_1,p_2,q \leq 1$, the coherence measures have the following relationships:
	
(B1) For any $0<\alpha_1<1$, $0\leq p_1,p_2 \leq \frac{1}{2}$ and $0\leq q \leq \frac{1}{2}$,
\begin{eqnarray}
&&C^{T}_{\frac{1}{2},1}(| \psi\rangle)\leq C^{T}_{\frac{1}{2},1}( |\varphi\rangle) \Leftrightarrow C^{T}_{\alpha_1}(| \psi\rangle)\leq C^{T}_{\alpha_1}( |\varphi\rangle)\nonumber\\
&&\Leftrightarrow C^{R}_{\alpha_1}(| \psi\rangle)\leq C^{R}_{\alpha_1}( |\varphi\rangle)  \Leftrightarrow C_{l_1}(| \psi\rangle)\leq C_{l_1}( |\varphi\rangle);\nonumber
\end{eqnarray}
	
(B2) For any $0<\alpha_1<1$, $ \frac{1}{2} \leq p_1,p_2 \leq 1$ and $\frac{1}{2} \leq q \leq 1$,
\begin{eqnarray}
&&C^{T}_{\frac{1}{2},1}(| \psi\rangle)\geq C^{T}_{\frac{1}{2},1}( |\varphi\rangle) \Leftrightarrow C^{T}_{\alpha_1}(| \psi\rangle)\geq C^{T}_{\alpha_1}( |\varphi\rangle) \nonumber\\
&&\Leftrightarrow C^{R}_{\alpha_1}(| \psi\rangle)\geq C^{R}_{\alpha_1}( |\varphi\rangle)  \Leftrightarrow C_{l_1}(| \psi\rangle)\geq C_{l_1}( |\varphi\rangle);\nonumber
\end{eqnarray}
	
(B3) For any $0<\alpha_1<1$, $0\leq p_1,p_2 \leq \frac{1}{2}$ and $\frac{1}{2} \leq q \leq 1$,
\begin{eqnarray}
&&C^{T}_{\frac{1}{2},1}(| \psi\rangle)\geq C^{T}_{\frac{1}{2},1}( |\varphi\rangle) \Leftrightarrow C^{T}_{\alpha_1}(| \psi\rangle)\leq C^{T}_{\alpha_1}( |\varphi\rangle)\nonumber\\
&&\Leftrightarrow C^{R}_{\alpha_1}(| \psi\rangle)\leq C^{R}_{\alpha_1}( |\varphi\rangle)  \Leftrightarrow C_{l_1}(| \psi\rangle)\leq C_{l_1}( |\varphi\rangle);\nonumber
\end{eqnarray}
	
(B4) For any $0<\alpha_1<1$, $ \frac{1}{2} \leq p_1,p_2 \leq 1$ and $0\leq q \leq \frac{1}{2}$,
\begin{eqnarray}
&&C^{T}_{\frac{1}{2},1}(| \psi\rangle)\leq C^{T}_{\frac{1}{2},1}( |\varphi\rangle) \Leftrightarrow C^{T}_{\alpha_1}(| \psi\rangle)\geq C^{T}_{\alpha_1}( |\varphi\rangle) \nonumber\\
&&\Leftrightarrow C^{R}_{\alpha_1}(| \psi\rangle)\geq C^{R}_{\alpha_1}( |\varphi\rangle)  \Leftrightarrow C_{l_1}(| \psi\rangle)\geq C_{l_1}( |\varphi\rangle).\nonumber
\end{eqnarray}
\end{theorem}

It is worthy noting that the Theorem \ref{Th1} is only valid for single-qubit pure states. The following example shows there could be different orderings for higher-dimensional systems.

\begin{example}
Consider the following two pure states in three-dimensional systems \cite{Liu16,Zhang},
\begin{eqnarray}
&&|\psi_{1}\rangle=\sqrt{\frac{12}{25}}|0\rangle+\sqrt{\frac{12}{25}}| 1\rangle+\sqrt{\frac{12}{25}}|2\rangle,\nonumber\\
&&|\psi_{2}\rangle=\sqrt{\frac{7}{10}}|0\rangle+\sqrt{\frac{2}{10}}| 1\rangle+\sqrt{\frac{1}{10}}|2\rangle.\nonumber
\end{eqnarray}
Let $\sigma=q_1|0\rangle \langle0|+ q_2|1\rangle \langle 1|+(1- q_1-q_2)|2\rangle \langle 2|$ with $q_1,q_2 \in [0,1]$ be the incoherent state. It is easy to calculate that $C_{l_{1}}(|\psi_{1}\rangle)=1.5143$, $C_{l_{1}}(|\psi_{2}\rangle)=1.5603$,
\begin{eqnarray}
C^{T}_{\frac{1}{2},1}(|\psi_{1}\rangle)&\geq&\min_{q_1, q_2}\{-\frac{2}{25}(11q_1+11 q_2-24)\}, \nonumber\\
C^{T}_{\frac{1}{2},1}(|\psi_{2}\rangle)&\geq&\min_{q_1, q_2}\{\frac{1}{5} (9-6q_1-q_2)\}, \nonumber\\	
C^{T}_{\alpha_1}(|\psi_{1}\rangle)&=&\frac{ (2^{\frac{2}{\alpha_1 }+1}+3^{\frac{1}{\alpha_1}}+1)^{\alpha_1 }-1}{25(\alpha_1 -1)},\nonumber\\
C^{T}_{\alpha_1}(|\psi_{2}\rangle)&=&\frac{(2^{\frac{1}{\alpha_1}}+7^{\frac{1}{\alpha_1} }+1)^{\alpha_1}-1}{10(\alpha_1 -1)},\nonumber\\
C^{R}_{\alpha_1}(|\psi_{1}\rangle)&=&\frac{\alpha_1}{\alpha_1 -1} \log \left(25^{-\frac{1}{\alpha_1}} (2^{\frac{2}{\alpha_1 }+1}+3^{\frac{1}{\alpha_1} }+1)\right), \nonumber\\
C^{R}_{\alpha_1}(|\psi_{2}\rangle)&=&\frac{\alpha_1}{\alpha_1 -1}  \log \left(10^{-\frac{1}{\alpha_1}}(2^{\frac{1}{\alpha_1} }+7^{\frac{1}{\alpha_1} }+1)\right). \nonumber
\end{eqnarray}
Set $\Delta C=C(|\psi_{2}\rangle)-C(|\psi_{1}\rangle)$. For any $0 \leq q_1,q_2 \leq 1$, there exist $q_1^{\ast }\in q_1$ and $q_2^{\ast } \in q_2$ such that
\begin{eqnarray}
C^{T}_{\frac{1}{2},1}(|\psi_{1}\rangle)
&=&-\frac{2}{25}(11q^{\ast }_1+11 q^{\ast }_2-24), \nonumber\\
C^{T}_{\frac{1}{2},1}(|\psi_{2}\rangle)
&=&\frac{1}{5} (9-6q^{\ast }_1-q^{\ast }_2).\nonumber
\end{eqnarray}
Then we have
\begin{eqnarray}
\Delta C^{T}_{\frac{1}{2},1}&=&\frac{1}{25} (-8q^{\ast }_1+17q^{\ast }_2-3), \nonumber\\
\Delta C^{T}_{\alpha_1}&=&\frac{1}{\alpha_1 -1}\Big[(10^{-\frac{1}{\alpha_1}} (2^{\frac{1}{\alpha_1}}+7^{\frac{1}{\alpha_1}}+1))^{\alpha_1 }\nonumber\\
&&-(5^{-\frac{2}{\alpha_1}} (2^{\frac{\alpha_1 +2}{\alpha_1 }}+3^{\frac{1}{\alpha_1} }+1))^{\alpha_1 }\Big] ,\nonumber\\
\Delta C^{R}_{\alpha_1}&=&\frac{\alpha_1}{\alpha_1 -1}\log \frac{10^{-\frac{1}{\alpha_1}}(2^{\frac{1}{\alpha_1}}+7^{\frac{1}{\alpha_1} }+1)}{25^{-\frac{1}{\alpha_1}}(2^{\frac{2}{\alpha_1 }+1}+ 3^{\frac{1}{\alpha_1}}+1)}. \nonumber
\end{eqnarray}
As can be seen from Fig. \ref{Fig2}, when $q^{\ast }_1, q^{\ast }_2 \in [0,1]$ $\Delta C^{T}_{\frac{1}{2},1}$ is less than 0 at first and then greater than 0.
$C_{l_{1}}(|\psi_{1}\rangle)< C_{l_{1}}(|\psi_{2}\rangle)$, $C^{T}_{\alpha_1}(|\psi_{1}\rangle)> C^{T}_{\alpha_1}(|\psi_{2}\rangle)$ and $C^{R}_{\alpha_1}(|\psi_{1}\rangle)> C^{R}_{\alpha_1}(|\psi_{2}\rangle)$, see Fig. \ref{Fig3}.
Therefore, $C^{T}_{\frac{1}{2},1}$ and $C^{T}_{\alpha_1}, C^{R}_{\alpha_1}$ generate different ordering for single-qutrit pure states $|\psi_{1}\rangle$ and $|\psi_{2}\rangle$.
\begin{figure}
	\includegraphics[width=7cm]{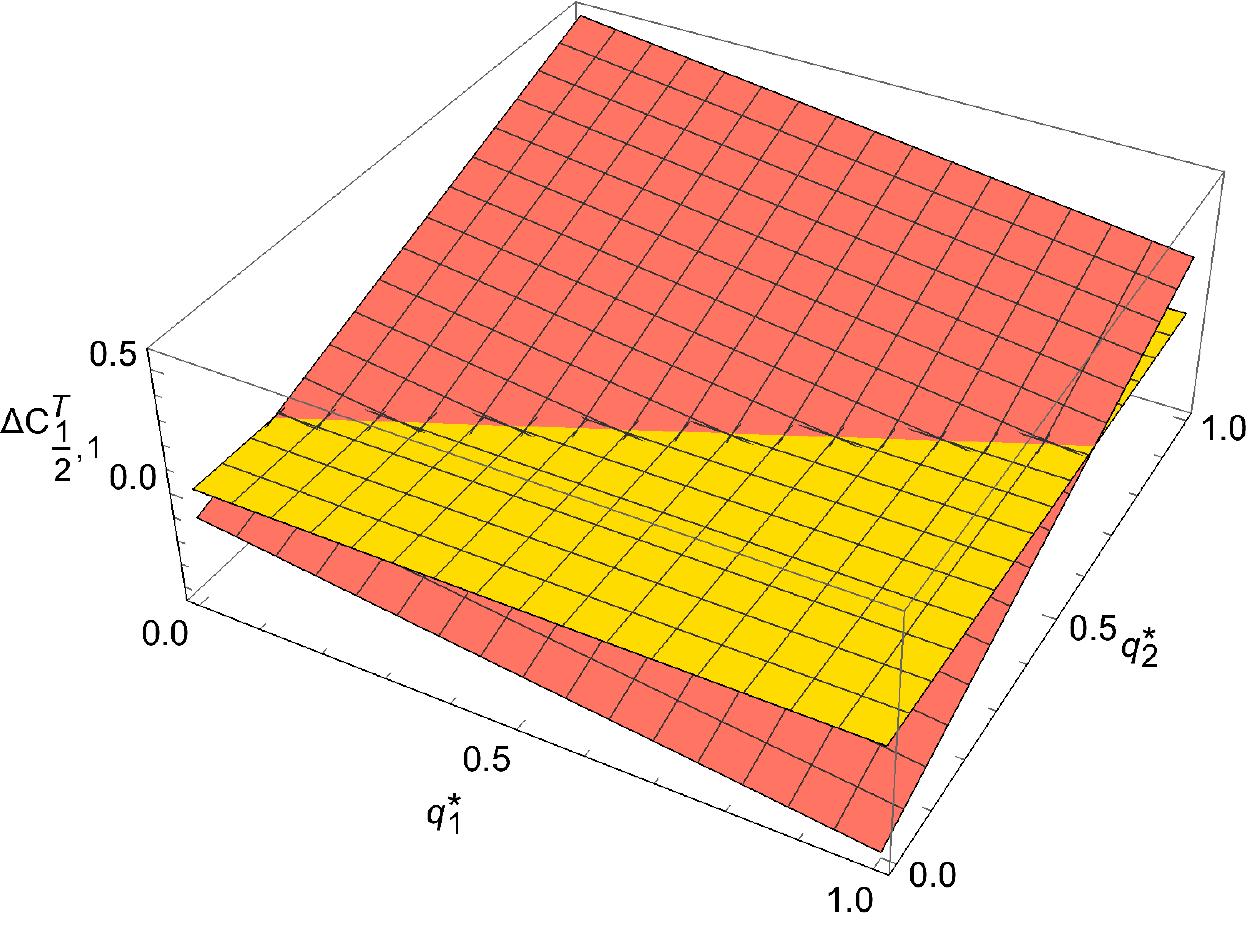}
	\caption{The pink surface is $\Delta C^{T}_{\frac{1}{2},1}$, and the yellow surface is the zero plane.}
	\label{Fig2}
\end{figure}
\begin{figure}
	\includegraphics[width=7cm]{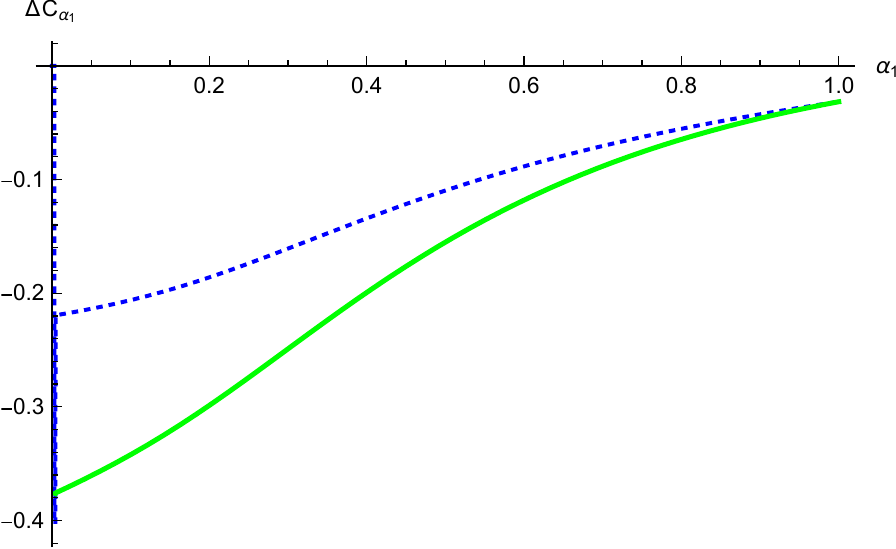}
	\caption{The blue dotted line is $\Delta C^{T}_{\alpha_1}$, and the green solid line is $\Delta C^{R}_{\alpha_1}$.}
	\label{Fig3}
\end{figure}
\end{example}

\subsection{Ordering states with $C^{T}_{\frac{1}{2},1}$, $C^{T}_{\frac{1}{2}}$,  $C^{R}_{\frac{1}{2}}$ and $C_{l_{1}}$ for single-qubit mixed states}
Any single-qubit state $\rho$ can be written as \cite{Liu16,Zhang},
\begin{equation}\label{Msrho}
\rho(t,z)=\left[ \begin {array}{cc} \frac{1+z}{2}&\frac{t}{2}\\ \noalign{\medskip}\frac{t}{2}&\frac{1-z}{2}\end {array} \right]
\end{equation}
with $t^2+z^2\leq 1$. $t^2+z^2=1$ if and only if $\rho(t,z)$ is a pure state.
Set $\sigma=\tilde{q}|0 \rangle \langle 0|+(1-\tilde{q})|1 \rangle \langle 1|$, where $0 \leq \tilde{q} \leq 1$. By substituting Eq.(\ref{Msrho}) into (\ref{Cl1}), (\ref{CT1}), (\ref{CR1}) and (\ref{Ctab2}), we get the coherence of $\rho(t,z)$ as follows,
\begin{eqnarray}
C^{T}_{\frac{1}{2},1}(\rho(t,z))&\geq&\min_{\tilde{q}}\{1+z-2\tilde{q}z-\sqrt{M^2_1-M^2_2}\},\nonumber\\
C^{T}_{\frac{1}{2}}(\rho(t,z))&=&-2(r^{\frac{1}{2}}-1) ,\nonumber\\
C^{R}_{\frac{1}{2}}(\rho(t,z))&=&-\log r ,\nonumber\\
C_{l_{1}}(\rho(t,z))&=&t, \nonumber
\end{eqnarray}
where
\begin{eqnarray}
M_1&=&(2\tilde{q}-1) z+1,\nonumber\\
M_2&=&\sqrt{4\tilde{q}z+z^2-2 z+1-4\tilde{q}(\tilde{q}-1)(t^2-1)},\nonumber\\
r&=&\sqrt{\frac{\sqrt{1-\sqrt{t^2+z^2}} \left(\sqrt{t^2+z^2}-z\right)}{2 \sqrt{2} \sqrt{t^2+z^2}}-N_1}\nonumber\\
&&+\sqrt{\frac{\left(\sqrt{t^2+z^2}+z\right) \sqrt{1-\sqrt{t^2+z^2}}}{2 \sqrt{2} \sqrt{t^2+z^2}}+N_2},\nonumber\\
N_1&=&\frac{\left(-\sqrt{t^2+z^2}-z\right) \sqrt{\sqrt{t^2+z^2}+1}}{2 \sqrt{2} \sqrt{t^2+z^2}},\nonumber\\
N_2&=&\frac{\left(\sqrt{t^2+z^2}-z\right) \sqrt{\sqrt{t^2+z^2}+1}}{2 \sqrt{2} \sqrt{t^2+z^2}}.\nonumber
\end{eqnarray}

For any $0 \leq \tilde{q} \leq 1$, suppose there exists $\tilde{q}^{\ast }\in \tilde{q}$ such that
\begin{eqnarray}
C^{T}_{\frac{1}{2},1}(\rho(t,z))
&=&1+z-2\tilde{q}^{\ast }z-\sqrt{\tilde{M}^2_1-\tilde{M}^2_2},
\end{eqnarray}
where $ \tilde{M}_2=\sqrt{4\tilde{q}^{\ast }z+z^2-2 z+1-4\tilde{q}^{\ast }(\tilde{q}^{\ast }-1)(t^2-1)}$ and $\tilde{M}_1=(2\tilde{q}^{\ast }-1) z+1.$
Consider the derivations of the coherence measures with respect to $t$, we have
\begin{eqnarray}
\frac{\partial C^{T}_{\frac{1}{2},1}}{\partial t}&=&-\frac{4 (\tilde{q}^{\ast }-1) \tilde{q}^{\ast } t}{\sqrt{\tilde{M}^2_1-\tilde{M}^2_2}} ,\nonumber\\
\frac{\partial C^{T}_{\frac{1}{2}}}{\partial t}&=&-r^{-\frac{1}{2}} \frac{\partial r}{\partial t} ,\nonumber\\
\frac{\partial C^{R}_{\frac{1}{2}}}{\partial t}&=&-\log r\frac{\partial r}{\partial t},\nonumber\\
\frac{\partial C_{l_1}}{\partial t}&=&1. \nonumber
\end{eqnarray}
Since $t^2+z^2<1$, assuming that $t^2+z^2=a$ and $a\in(0,1)$, we can get the analytical expression of $\frac{\partial r}{\partial t}$, as shown in Fig. \ref{Fig4}. The expression of $\frac{\partial r}{\partial t}$ is given in Appendix B. One sees that when $0\leq t < \sqrt{1-z^2}$, $\frac{\partial C^{T}_{\frac{1}{2},1}}{\partial t} \geq 0$, $\frac{\partial C^{T}_{\frac{1}{2}}}{\partial t}$, $\frac{\partial C^{R}_{\frac{1}{2}}}{\partial t} \geq 0$; and $\frac{\partial C^{T}_{\frac{1}{2},1}}{\partial t} \leq 0$, $\frac{\partial C^{T}_{2}}{\partial t}$ and $\frac{\partial C^{R}_{2}}{\partial t} \leq 0$ when $-\sqrt{1-z^2}< t\leq 0$.
\begin{figure}[tb]
	\includegraphics[width=7cm]{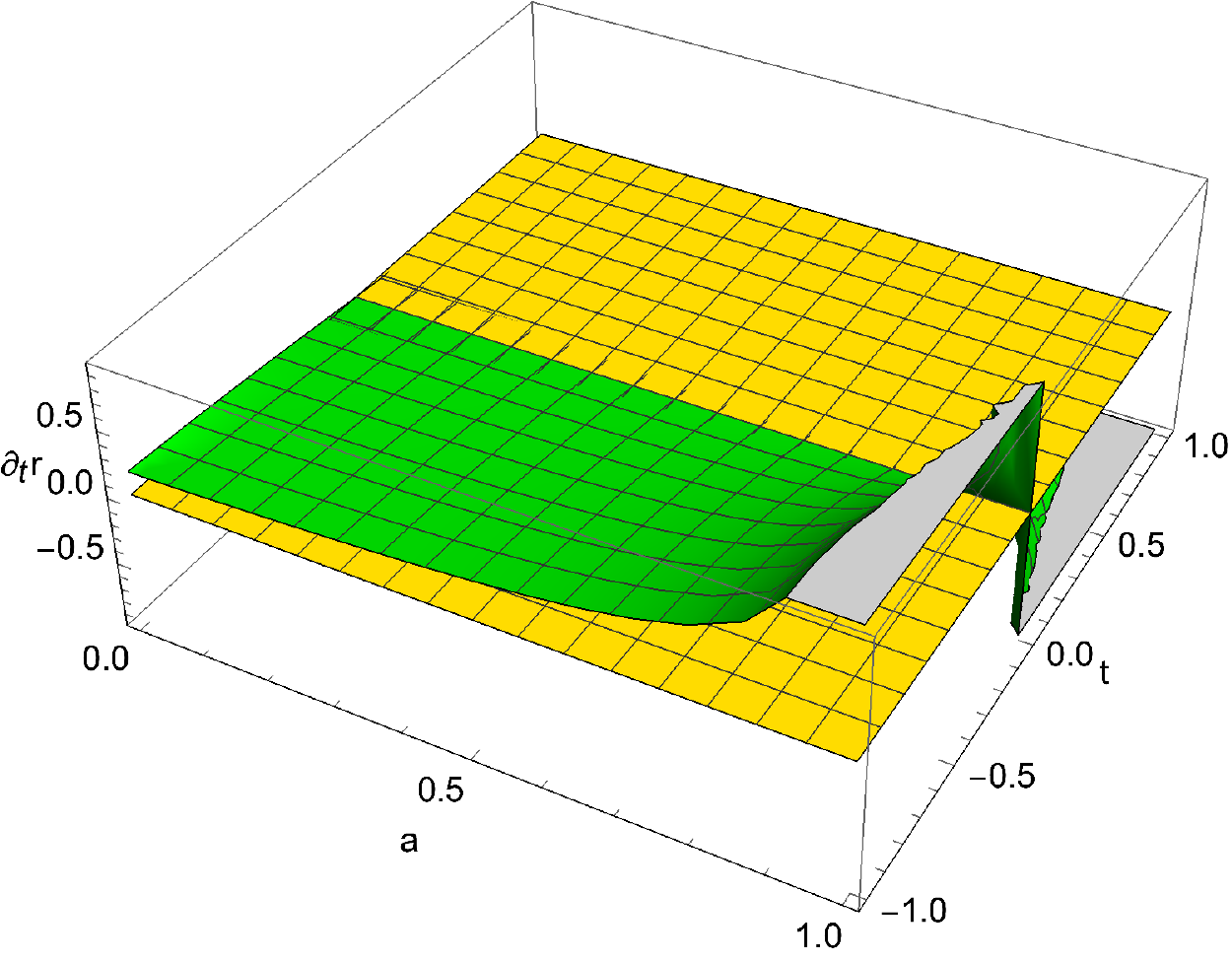}
	\caption{The green surface is $\frac{\partial r}{\partial t}$, and the yellow surface is zero plane.}
	\label{Fig4}
\end{figure}

Concerning the monotonicity of these coherence measures with respect to variable $t$, we have
(a) For any $-\sqrt{1-z^2}<t<\sqrt{1-z^2}$, $C_{l_{1}}$ is an increasing function.
(b) For any $-\sqrt{1-z^2}<t\leq 0$, $C^{T}_{\frac{1}{2},1}$, $C^{T}_{\frac{1}{2}}$ and $C^{R}_{\frac{1}{2}}$ are decreasing function, and for $-\sqrt{1-z^2}< t\leq 0$ they are increasing function. Without loss of generality, set $0\leq p_1,p_2 \leq 1$ and $t\leq 0$.
We have $C^{T}_{\frac{1}{2},1}(| \psi\rangle)\geq C^{T}_{\frac{1}{2},1}( |\varphi\rangle)$ if and only if $p_1\leq p_2$, $C^{T}_{\frac{1}{2}}(| \psi\rangle)\leq C^{T}_{\frac{1}{2}}( |\varphi\rangle)$ if and only if $p_1\leq p_2$, $C^{R}_{\frac{1}{2}}(| \psi\rangle)\leq C^{R}_{\frac{1}{2}}( |\varphi\rangle)$ if and only if $p_1\leq p_2$, and $p_1\leq p_2$ if and only if $C_{l_1}(| \psi\rangle)\leq C_{l_1}( |\varphi\rangle)$. Therefore, we have the following theorem.

\begin{theorem}
For any two qubit mixed states $\rho_{1}$ and $\rho_{2}$ of the form(\ref{Msrho}), and for any $t^2+z^2 <1$, $C^{T}_{\frac{1}{2}, 1}$, $C^{T}_{\frac{1}{2}}$, $C^{R}_{\frac{1}{2}}$ and $C_{l_{1}}$ satisfy the following relationships,
\begin{eqnarray}\label{Th21}
&&C^{T}_{\frac{1}{2}, 1}(\rho_{1})\leq C^{T}_{\frac{1}{2}, 1}( \rho_{2}) \Leftrightarrow C^{T}_{\frac{1}{2}}(\rho_{1})\leq C^{T}_{\frac{1}{2}}(\rho_{2}) \nonumber\\
&&\Leftrightarrow C^{R}_{\frac{1}{2}}(\rho_{1})\leq C^{R}_{\frac{1}{2}}(\rho_{1})  \Leftrightarrow C_{l_1}(\rho_{1})\leq C_{l_1}(\rho_{2});
\end{eqnarray}
or
\begin{eqnarray}\label{Th22}
&&C^{T}_{\frac{1}{2}, 1}(\rho_{1})\geq C^{T}_{\frac{1}{2}, 1}( \rho_{2}) \Leftrightarrow C^{T}_{\frac{1}{2}}(\rho_{1})\geq C^{T}_{\frac{1}{2}}(\rho_{2})\nonumber\\
&&\Leftrightarrow C^{R}_{\frac{1}{2}}(\rho_{1})\geq C^{R}_{\frac{1}{2}}(\rho_{2})  \Leftrightarrow C_{l_1}(\rho_{1})\leq C_{l_1}(\rho_{2}).
\end{eqnarray}
\end{theorem}

\begin{figure}[tb]
	\includegraphics[width=7cm]{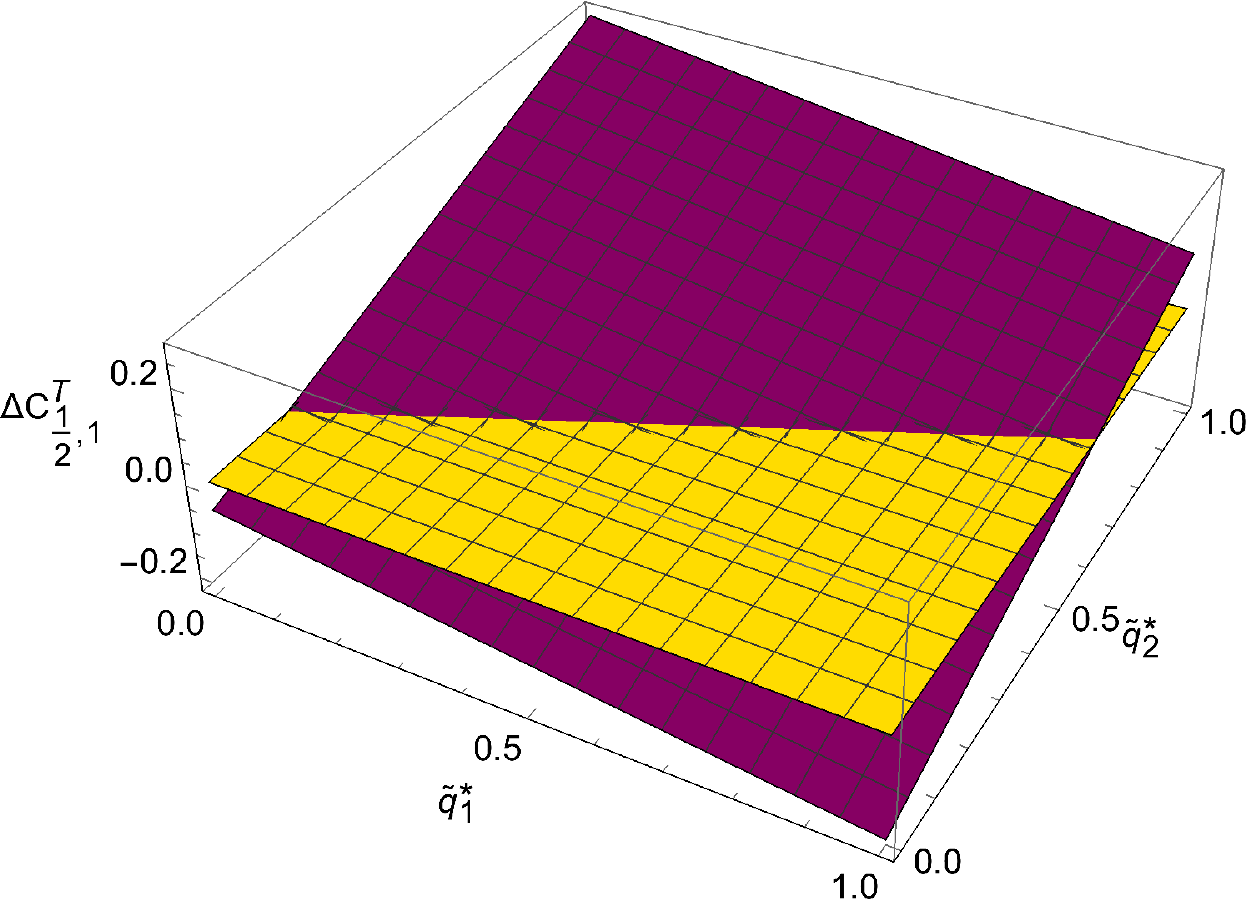}
	\caption{The purple surface is $\Delta C^{T}_{\frac{1}{2},1}$, and the yellow surface is the zero plane.}
	\label{Fig5}
\end{figure}

Theorem \ref{Th2} holds for qubit mixed states. For higher-dimensional systems there could be different orderings.
\begin{example}\label{exam2}
Consider the following two mixed states in three-dimensional systems,
\begin{eqnarray}
\rho_{1}=\left(
\begin{array}{ccc}
	\frac{6}{25} & \frac{6}{25} & \frac{\sqrt{3}}{25} \\
	\frac{6}{25} & \frac{6}{25} & \frac{\sqrt{3}}{25} \\
	\frac{\sqrt{3}}{25} & \frac{\sqrt{3}}{25} & \frac{1}{50} \\
\end{array}
\right),~~\nonumber
\rho_{2}=\left(
\begin{array}{ccc}
	\frac{7}{20} & \frac{\sqrt{\frac{7}{2}}}{10} & \frac{\sqrt{7}}{20} \\
	\frac{\sqrt{\frac{7}{2}}}{10} & \frac{1}{10} & \frac{1}{10 \sqrt{2}} \\
	\frac{\sqrt{7}}{20} & \frac{1}{10 \sqrt{2}} & \frac{1}{20} \\
\end{array}
\right).\nonumber
\end{eqnarray}
Let $\sigma=\tilde{q}_1|0\rangle \langle0|+ \tilde{q}_2|1\rangle \langle 1|+(1- \tilde{q}_1-\tilde{q}_2)|2\rangle \langle 2|$ with $\tilde{q}_1, \tilde{q}_2 \in [0,1]$.
It is easy to calculate that $C_{l_{1}}(\rho_{1})=0.7571$, $C_{l_{1}}(\rho_{2})=0.7802$, $C^{T}_{\frac{1}{2}}(\rho_{1})=1.0383$, $C^{T}_{\frac{1}{2}}(\rho_{2})=0.9608$, $C^{R}_{\frac{1}{2}}(\rho_{1})=1.4645$, $C^{R}_{\frac{1}{2}}(\rho_{2})=1.3093$,
\begin{eqnarray}
C^{T}_{\frac{1}{2},1}(\rho_{1})&\geq&\min_{\tilde{q}_1, \tilde{q}_2}\{\frac{1}{25} (49-11\tilde{q}_1-11\tilde{q}_2)\}, \nonumber\\
C^{T}_{\frac{1}{2},1}(\rho_{2})&\geq&\min_{\tilde{q}_1, \tilde{q}_2}\{\frac{1}{10} (19-6\tilde{q}_1-\tilde{q}_2)\}. \nonumber
\end{eqnarray}
Set $\Delta C=C(\rho_{2})-C(\rho_{1})$. For any $0 \leq \tilde{q}_1,\tilde{q}_2 \leq 1$, suppose there exist $\tilde{q}_1^{\ast}\in \tilde{q}_1$ and $\tilde{q}_2^{\ast }\in \tilde{q}_2$ such that
\begin{eqnarray}
C^{T}_{\frac{1}{2},1}(\rho_{1})
&=&\frac{1}{25} (49-11\tilde{q}^{\ast}_1-11\tilde{q}^{\ast}_2),\nonumber\\
C^{T}_{\frac{1}{2},1}(\rho_{2})
&=&\frac{1}{10} (19-6\tilde{q}^{\ast}_1-\tilde{q}^{\ast}_2).\nonumber
\end{eqnarray}
Then we have
\begin{eqnarray}
\Delta C^{T}_{\frac{1}{2},1}&=&\frac{1}{50} (-8\tilde{q}^{\ast}_1+17 \tilde{q}^{\ast}_2-3).\nonumber
\end{eqnarray}
It is clear that $C_{l_{1}}(\rho_{1})< C_{l_{1}}(\rho_{2})$, $C^{T}_{\frac{1}{2}}(\rho_{1})> C^{T}_{\frac{1}{2}}(\rho_{2})$ and $C^{R}_{\frac{1}{2}}(\rho_{1})> C^{R}_{\frac{1}{2}}(\rho_{2})$.
As can be seen from Fig. \ref{Fig5}, when $\tilde{q}^{\ast}_1, \tilde{q}^{\ast}_2 \in [0,1]$, $\Delta C^{T}_{\frac{1}{2},1}$ varies from less than 0 to greater than 0.
Hence, $C^{T}_{\frac{1}{2},1}$ and $C^{T}_{\frac{1}{2}}$, $C^{R}_{\frac{1}{2}}$ generate different ordering for the qutrit mixed states $\rho_{1}$ and $\rho_{2}$.
\end{example}

\section{Conclusion}
In conclusion, we have proposed a quantum coherence measure based on the perspective mapping of function $\ln_{\alpha} x:=\frac{x^{\alpha}-1}{\alpha}$, that is, Tsallis relative operator $(\alpha, \beta)$-entropy. It has been demonstrated that this coherence measure meets all the necessary criteria for satisfactory coherence measures, especially with the property of strong monotonicity. We have further investigated the ordering of the Tsallis relative operator $(\alpha, \beta)$-entropy of coherence, Tsallis relative $\alpha$-entropies of coherence, R\'{e}nyi $\alpha$-entropy of coherence and $l_{1}$ norm of coherence for single qubit states. When $\alpha=\frac{1}{2}$ and $\beta=1$, we have proved that under certain conditions, the Tsallis relative operator $(\alpha, \beta)$-entropy of choerence, Tsallis relative $\alpha$-entropies of coherence, R\'{e}nyi $\alpha$-entropy of coherence and $l_{1}$ norm of coherence give the same ordering for pure qubit states. The results are extended to the case of qubit mixed states. Our results provide a new method for defining a new good coherence measure and a new idea for further study of quantum coherence.

\section*{ACKNOWLEDGMENTS}
This work is supported by the National Natural Science Foundation of China (NSFC) under Grants 12075159, 12171044 and 12175147; Beijing Natural Science Foundation (Grant No. Z190005); the Academician Innovation Platform of Hainan Province.

\section*{Appendix}
\subsection{proof of T1-T4}
(T1) (monotonicity)

For positive operators $\rho$, $\sigma$ and $\tau$ with $\sigma \leq \tau$, and any real numbers $\alpha\in (0,1]$ and $\beta \in (0,1]$, we have
\begin{eqnarray}
\rho^{-\frac{ \beta}{2}}\sigma \rho^{-\frac{ \beta}{2}} &\leq& \rho^{-\frac{ \beta}{2}} \tau \rho^{-\frac{ \beta}{2}} \nonumber \\
(\rho^{-\frac{ \beta}{2}}\sigma \rho^{-\frac{ \beta}{2}}) ^{\alpha} &\leq& (\rho^{-\frac{ \beta}{2}} \tau \rho^{-\frac{ \beta}{2}})^{\alpha} \nonumber \\
(\rho^{-\frac{ \beta}{2}}\sigma \rho^{-\frac{ \beta}{2}}) ^{\alpha} -I &\leq& (\rho^{-\frac{ \beta}{2}} \tau \rho^{-\frac{ \beta}{2}})^{\alpha}-I \nonumber \\
\frac{\rho^{\frac{ \beta}{2}}[(\rho^{-\frac{ \beta}{2}}\sigma \rho^{-\frac{ \beta}{2}}) ^{\alpha} -I] \rho^{\frac{ \beta}{2}}}{\alpha}&\leq& \frac{\rho^{\frac{ \beta}{2}}[(\rho^{-\frac{ \beta}{2}} \tau \rho^{-\frac{ \beta}{2}})^{\alpha}-I ] \rho^{\frac{ \beta}{2}}}{\alpha}. \nonumber
\end{eqnarray}

(T2) (superadditivity)

For $\rho_1 \leq \rho_2$ and $\sigma_1 \leq \sigma_2$, assume that both $\sigma_1$ and $\sigma_2$ are invertible. Set $X=\rho_1^{\frac{\beta}{2}} (\rho_1 +\rho_2)^{-\frac{\beta}{2}} $, $X^{*}=(\rho_1 +\rho_2)^{-\frac{\beta}{2}} \rho_1^{\frac{\beta}{2}}$, $Y=\rho_2^{\frac{\beta}{2}} (\rho_1 +\rho_2)^{-\frac{\beta}{2}} $ and $Y^{*}=(\rho_1 +\rho_2)^{-\frac{\beta}{2}} \rho_2^{\frac{\beta}{2}}$. Since $X^{*}X +Y^{*}Y = I_H$, where $I_H$ is an identity operator in $\mathcal{B}(H)$, $\mathcal{B}(H)$ is a semi-algebra of all bounded linear operators
on Hilbert space $\mathcal{H}$ to $\mathcal{H}$. Then we obtain
\begin{eqnarray}
&&T_{\alpha, \beta}(\rho_1+\rho_2||\sigma_1+\sigma_2) \nonumber \\
=&&(\rho_1 +\rho_2)^{\frac{\beta}{2}} \ln_{\alpha}[(\rho_1 +\rho_2)^{-\frac{\beta}{2}} (\sigma_1 +\sigma_2) (\rho_1 +\rho_2)^{-\frac{\beta}{2}}]\times \nonumber\\
&& (\rho_1 +\rho_2)^{\frac{\beta}{2}} \nonumber\\
=&&(\rho_1 +\rho_2)^{\frac{\beta}{2}} \ln_{\alpha}(X^{*} \rho^{-\frac{\beta}{2}}_1 \sigma_1 \rho^{-\frac{\beta}{2}}_1 X +Y^{*} \rho^{-\frac{\beta}{2}}_2 \sigma_2 \rho^{-\frac{\beta}{2}}_2 Y )\times \nonumber\\
&& (\rho_1 +\rho_2)^{\frac{\beta}{2}} \nonumber \\
\geq&&(\rho_1 +\rho_2)^{\frac{\beta}{2}} [X^{*} (\ln_{\alpha}( \rho^{-\frac{\beta}{2}}_1 \sigma_1 \rho^{-\frac{\beta}{2}}_1))X \nonumber\\
&&+Y^{*}\ln_{\alpha} (\rho^{-\frac{\beta}{2}}_2 \sigma_2 \rho^{-\frac{\beta}{2}}_2) Y )](\rho_1 +\rho_2)^{\frac{\beta}{2}} \nonumber \\
=&&\rho_1^{\frac{\beta}{2}} \ln_{\alpha}( \rho^{-\frac{\beta}{2}}_1 \sigma_1 \rho^{-\frac{\beta}{2}}_1) \rho_1^{\frac{\beta}{2}} +\rho_2^{\frac{\beta}{2}} \ln_{\alpha}( \rho^{-\frac{\beta}{2}}_2 \sigma_2 \rho^{-\frac{\beta}{2}}_2) \rho_2^{\frac{\beta}{2}}  \nonumber \\
=&&T_{\alpha, \beta}(\rho_1||\sigma_1)+T_q(\rho_2||\sigma_2), \nonumber
\end{eqnarray}
where the inequality follows from the Theorem 1.9 in \cite{OP}.

(T4) For any unitary operator $U$, we have
\begin{eqnarray}
&&T_{\alpha, \beta}(U \rho U^{\dagger} || U \sigma U^{\dagger }) \nonumber \\
&&=(U \rho U^{\dagger})^{\frac{\beta}{2}} \ln_{\alpha}((U \rho U^{\dagger})^{-\frac{\beta}{2}} (U \sigma U^{\dagger }) (U \rho U^{\dagger})^{-\frac{\beta}{2}}) (U \rho U^{\dagger})^{\frac{\beta}{2}} \nonumber\\
&&=U \rho^{\frac{\beta}{2}} U^{\dagger} \ln_{\alpha}(U \rho^{-\frac{\beta}{2}} U^{\dagger} U \sigma U^{\dagger } U \rho^{-\frac{\beta}{2}} U^{\dagger}) U \rho^{\frac{\beta}{2}} U^{\dagger} \nonumber\\
&&=U \rho^{\frac{\beta}{2}} U^{\dagger} U \ln_{\alpha}( \rho^{-\frac{\beta}{2}} \sigma \rho^{-\frac{\beta}{2}} ) U^{\dagger} U \rho^{\frac{\beta}{2}} U^{\dagger} \nonumber\\
&&=U \rho^{\frac{\beta}{2}} \ln_{\alpha}( \rho^{-\frac{\beta}{2}} \sigma \rho^{-\frac{\beta}{2}} ) \rho^{\frac{\beta}{2}} U^{\dagger} \nonumber\\
&&=UT_{\alpha, \beta}(\rho||\sigma)U^{\dagger}. \nonumber
\end{eqnarray}

(T5) Assume $\rho$ is invertible, then so does $\Phi(\rho)$.
Define $\Phi_\rho (X)= \Phi (\rho)^{-\frac{\beta}{2}} \Phi ( \rho^{\frac{\beta}{2}} X \rho^{\frac{\beta}{2}} ) \Phi (\rho)^{-\frac{\beta}{2}} $.
So $\Phi_\rho$ is a normalized positive
linear map. Consequently,
\begin{eqnarray}
&&\Phi(T_{\alpha, \beta}(\rho||\sigma)) \nonumber \\
&&=\Phi (\rho^{\frac{\beta}{2}} \ln_{\alpha}( \rho^{-\frac{\beta}{2}} \sigma \rho^{-\frac{\beta}{2}} ) \rho^{\frac{\beta}{2}} ) \nonumber\\
&&=\Phi (\rho)^{\frac{\beta}{2}} \Phi_\rho( \ln_{\alpha}( \rho^{-\frac{\beta}{2}} \sigma \rho^{-\frac{\beta}{2}} ) ) \Phi (\rho)^{\frac{\beta}{2}}  \nonumber\\
&&\leq  \Phi (\rho)^{\frac{\beta}{2}} \ln_{\alpha} (\Phi_\rho( \rho^{-\frac{\beta}{2}} \sigma \rho^{-\frac{\beta}{2}} ) ) \Phi (\rho)^{\frac{\beta}{2}} \nonumber\\
&&= \Phi (\rho)^{\frac{\beta}{2}} \ln_{\alpha} [\Phi (\rho)^{-\frac{\beta}{2}} \Phi ( \rho^{\frac{\beta}{2}} \rho^{-\frac{\beta}{2}} \sigma \rho^{-\frac{\beta}{2}} \rho^{\frac{\beta}{2}} ) \Phi (\rho)^{-\frac{\beta}{2}}]  \Phi (\rho)^{\frac{\beta}{2}}\nonumber \\
&&= \Phi (\rho)^{\frac{\beta}{2}} \ln_{\alpha} [\Phi (\rho)^{-\frac{\beta}{2}} \Phi ( \sigma) \Phi (\rho)^{-\frac{\beta}{2}}]  \Phi (\rho)^{\frac{\beta}{2}}\nonumber\\
&&= T_{\alpha, \beta}(\Phi(\rho)||\Phi(\sigma)), \nonumber
\end{eqnarray}
where the inequality is due to the Davis-Choi-Jensen's inequality \cite{OP}: $\Phi_\rho (F(X)) \leq F(\Phi_\rho(X))$ for every operator concave function $F$ on $(0,\infty)$.

\subsection{Proof of Lemma \ref{lem2}}
According to the property (T5) and the Jensen's inequality one gets $\Phi[\rho^{\beta}] \leq \Phi(\rho)^{\beta}$ and
\begin{eqnarray}
&&\Phi[\rho^{\frac{\beta}{2}}(\rho^{-\frac{\beta}{2}}\sigma \rho^{-\frac{\beta}{2}})^{\alpha}\rho^{\frac{\beta}{2}}]\nonumber\\
&&\leq \Phi(\rho)^{\frac{\beta}{2}}(\Phi(\rho)^{-\frac{\beta}{2}}\Phi(\sigma) \Phi( \rho)^{-\frac{\beta}{2}})^{\alpha}\Phi(\rho)^{\frac{\beta}{2}}.\label{qaz32}
\end{eqnarray}
For any CPTP map $\Phi$, we have
\begin{eqnarray}
&&\mathrm{Tr}\Big[\Phi \Big(\rho^{\frac{\beta}{2}}(\rho^{-\frac{\beta}{2}}\sigma \rho^{-\frac{\beta}{2}})^{\alpha}\rho^{\frac{\beta}{2}}\Big)\Big]\nonumber\\
&&=\mathrm{Tr}[\rho^{\frac{\beta}{2}}(\rho^{-\frac{\beta}{2}}\sigma \rho^{-\frac{\beta}{2}})^{\alpha}\rho^{\frac{\beta}{2}}]\label{q32}
\end{eqnarray}
and
\begin{eqnarray}
&&\mathrm{Tr}\left[\Phi(\rho)^{\frac{\beta}{2}}(\Phi(\rho)^{-\frac{\beta}{2}}\Phi(\sigma) \Phi(\rho)^{-\frac{\beta}{2}})^{\alpha}\Phi(\rho)^{\frac{\beta}{2}}\right] \nonumber\\
&&=f(\Phi(\rho),\Phi(\sigma)).\label{q33}
\end{eqnarray}
According to (\ref {qaz32}), (\ref {q32}) and (\ref {q33}), we get $f(\rho,\sigma) \leq f(\Phi(\rho),\Phi(\sigma))$.

\subsection{Proof of Lemma \ref{SSS}}
Any TPCP map can be achieved by unitary operations and local projection measurements on the composite system \cite{sic,Guo2020}. Suppose the system $H$ is of interest to us and $A$ is an auxiliary system. For a TPCP map $\Phi :=\left\{ K_{n}:\sum_{n}{K}_{n}^{\dagger }{K}_{n}= \mathcal{I}_{H}\right\}$, we can always find a unitary operation $U_{HA}$ and a set of projectors
$\left\{ \Pi_{n}^{A}=\left\vert n \right\rangle_{A}\left\langle n\right\vert \right\} $ such that
\begin{eqnarray}
&&K_{n}\rho _{H}K_{n}^{\dagger }\otimes \Pi _{n}^{A}\nonumber\\
=&&\left( \mathcal{I}_{H}\otimes \Pi _{n}^{A}\right) U_{HA}\left( \rho_{H}\otimes \Pi _{0}^{A}\right) U_{HA}^{\dagger }\left( \mathcal{I}_{H}\otimes \Pi _{n}^{A}\right). \label{eq}
\end{eqnarray}
According to Lemma \ref{Lm1} and the property (T4), for any states $\rho _{H}$ and $\sigma _{H}$ we have
\begin{eqnarray}
f\left(\rho_{H},\sigma_{H}\right)
=f\left(U_{HA}\left( \rho _{H}\otimes \Pi _{0}^{A}\right)
U_{HA}^{\dagger},U_{HA}\left(\sigma_{H}\otimes \Pi _{0}^{A}\right)
U_{HA}^{\dagger}\right).\nonumber
\end{eqnarray}
Denote $\rho _{Hf}=\Phi_{HA}
\left[ U_{HA}\left( \rho_{H}\otimes \Pi _{0}^{A}\right) U_{HA}^{\dagger }
\right] $ and $\sigma _{Hf}=\Phi_{HA}\left[ U_{HA}\left( \sigma _{H}\otimes
\Pi _{0}^{A}\right) U_{HA}^{\dagger }\right]$.
Due to Lemma \ref{lem2}, we obtain
\begin{equation}\label{eq1r}
f\left( \rho _{H},\sigma_{H}\right) \leq f\left( \rho _{Hf},\sigma _{Hf}\right) .
\end{equation}

Let the TPCP map be given by  $\Phi_{HA}:=\left\{ \mathcal{I}_{H}\otimes \Pi_{n}^{A}\right\}$. According to (\ref{eq}), $\rho _{Hf}$ and $\sigma _{Hf}$ in (\ref{eq1r}) can be replaced by
\begin{eqnarray}
\rho _{Hf}\rightarrow \tilde{\rho}_{Hf}=\sum_{n}K_{n}\rho _{H}K_{n}^{\dagger }\otimes \Pi _{n}^{A}
\end{eqnarray}
and
\begin{eqnarray}
\sigma _{Hf}\rightarrow \tilde{\sigma}_{Hf}=\sum\limits_{n}K_{n}\sigma_{H}K_{n}^{\dagger }\otimes \Pi _{n}^{A},
\end{eqnarray}
respectively.
Thus, we have
\begin{eqnarray}
f\left( \rho _{H},\delta _{H}\right) &\leq& f\left( \tilde{\rho}_{H_f},\tilde{\sigma}_{H_f}\right) \nonumber\\
&=& \sum_{n}f\left( K_{n}\rho_{H}K_{n}^{\dagger }\otimes \Pi _{n}^{A}, K_{n}\sigma _{H}K_{n}^{\dagger}\otimes \Pi _{n}^{A}\right) \nonumber\\
&=& \sum_{n}f\left( K_{n}\rho_{H}K_{n}^{\dagger }, K_{n}\sigma _{H}K_{n}^{\dagger }\right)\nonumber\\
&=& \sum_{n}p_{n}^{\gamma}q_{n}^{1-\gamma}f\left( \rho _{n},\sigma _{n}\right),\nonumber
\end{eqnarray}
which comletes the proof.

\begin{widetext}
\subsection{The expression of $\frac{\partial r}{\partial t}$}
\begin{eqnarray}
\frac{\partial r}{\partial t}&=&\frac{1}{2^{\frac{11}{4}}a^{\frac{3}{2}}\sqrt{1-a}}
\Bigg[\frac{\left(\sqrt{1-\sqrt{a}}-\sqrt{\sqrt{a}+1}\right) t^3}{\sqrt{\frac{\left(\sqrt{\sqrt{a}+1}-\sqrt{1-\sqrt{a}}\right) \sqrt{a-t^2}}{\sqrt{a}}+\sqrt{1-\sqrt{a}}+\sqrt{\sqrt{a}+1}}}+\frac{\left(\sqrt{1-\sqrt{a}}-\sqrt{\sqrt{a}+1}\right) t^3}{\sqrt{\frac{\left(\sqrt{1-\sqrt{a}}-\sqrt{\sqrt{a}+1}\right) \sqrt{a-t^2}}{\sqrt{a}}+\sqrt{1-\sqrt{a}}+\sqrt{\sqrt{a}+1}}}\nonumber\\
&&+\frac{t \sqrt{a-t^2} \Big((\sqrt{1-\sqrt{a}}-\sqrt{\sqrt{a}+1}) \sqrt{a-t^2}-2 \sqrt{1-\sqrt{a}}-(\sqrt{1-\sqrt{a}}+\sqrt{\sqrt{a}+1}) \sqrt{a}+2 \sqrt{\sqrt{a}+1}\Big)}{\sqrt{\frac{(\sqrt{\sqrt{a}+1}-\sqrt{1-\sqrt{a}}) \sqrt{a-t^2}}{\sqrt{a}}+\sqrt{1-\sqrt{a}}+\sqrt{\sqrt{a}+1}}}\nonumber\\
&&+\frac{t \sqrt{a-t^2} \left((\sqrt{1-\sqrt{a}}-\sqrt{\sqrt{a}+1}) \sqrt{a-t^2}+2 (\sqrt{1-\sqrt{a}}-\sqrt{\sqrt{a}+1})+\sqrt{a} (\sqrt{1-\sqrt{a}}+\sqrt{\sqrt{a}+1})\right)}{\sqrt{\frac{\left(\sqrt{1-\sqrt{a}}-\sqrt{\sqrt{a}+1}\right) \sqrt{a-t^2}}{\sqrt{a}}+\sqrt{1-\sqrt{a}}+\sqrt{\sqrt{a}+1}}}\Bigg].\nonumber
\end{eqnarray}
\end{widetext}

\end{document}